\documentclass[aps,twocolumn,groupedaddress,showpacs,nofootinbib]{revtex4}

\usepackage{graphicx}% Include figure files
\usepackage{hyperref}
\usepackage{dcolumn}% Align table columns on decimal point
\usepackage{bm}% bold math
\usepackage{epsfig,amsbsy,bm,amssymb,amsmath}
\usepackage{xcolor}

\newcommand{\hs}{\hspace*}
\newcommand{\vs}{\vspace*}

\newcommand{\eref}[1] {(\ref{#1})}
\newcommand{\Eref}[1] {Eq.~(\ref{#1})}
\newcommand{\Fref}[1] {Fig. \ref{#1}}

\newcommand{\nn}{\nonumber}
\newcommand{\be}{\begin{equation}}
\newcommand{\ee}{\end{equation}}
\newcommand{\br}{\begin{eqnarray*}}
\newcommand{\er}{\end{eqnarray*}}
\newcommand{\ba}{\begin{eqnarray}}
\newcommand{\ea}{\end{eqnarray}}
\newcommand{\bp}{\begin{minipage}}
\newcommand{\ep}{\end{minipage}}
\newcommand{\bt}{\begin{tabular}}
\newcommand{\et}{\end{tabular}}

\newcommand{\ms}{\vspace*{-5mm}}

  \newcommand{\ve}{\varepsilon}

\newcommand{\N}{N$_2$~}

\begin{document}
\bibliographystyle{apsrev}

\title{Shape resonances in photoionization cross sections and time
  delay}

\author{Anatoli~S. Kheifets}
%\email{A.Kheifets@anu.edu.au}
\author{Stephen Catsamas}

\affiliation{$^{1}$Research School of Physics, The Australian National
  University, Canberra ACT 2601, Australia}

 \date{\today}


\begin{abstract}
Shape resonances in photoionization of atoms and molecules arise from
a particular geometry of the ionic potential which traps the receding
photoelectron in a quasi-bound state in a particular partial
wave. This mechanism allows us to connect the photoionization cross
section in the resonant region with the photoelectron scattering phase
in this partial wave by a simple formula $\sigma \propto
\sin^2\delta_\ell$. Due to this relation, the phase $\delta_\ell$ can
be extracted from an experimentally known cross section and then
converted to the photoelectron group delay (Wigner time delay)
$\tau_{\rm W} = \partial \delta_\ell/\partial E$ which is measurable
by recently developed laser interferometric techniques. Such a direct
connection of the photoionization cross section and the time delay is
a fundamental property of shape resonances which provides a
comprehensive test of novel measurements against a large body of
older synchrotron data.
 \end{abstract}

\maketitle
%\end{document} %stop

Shape resonances (SR's) affect profoundly numerous phenomena in
physics, chemistry and biology (see Introduction of \cite{Nandi2020}
for several fascinating examples).  SR's have long been intensely studied
in electron-molecule scattering \cite{Bardsley1968} and molecular
photoionization \cite{Dehmer1988}. Similar resonant features are
observed in electron-atom scattering \cite{Shimamura2012} and atomic
photoionization \cite{Rau1968,RevModPhys.40.441,Connerade1986}.
Formation of SR's is well understood
\cite{Bardsley1968,Carlson1983IEEE,Dehmer1988,Child1996,Shimamura2012}.
SR's are associated with the shape of some effective potential in an
open channel, normally a combination of short-range attractive and
long-range repulsive potentials. Such a combination forms a barrier
holding a large portion of the electron wave function while the
remaining part of this wave function leaks out.  This normally occurs
at energies above and usually close to the threshold of that open
channel and is typically associated with a large photoelectron angular
momentum $\ell \geq 2$.  A common property of SR's is that they can be
turned into bound states by a slight modification of the target
Hamiltonian \cite{Chrysos1998,Horacek2019}. In molecules, SR's can be
 associated with anti-bonding vacant orbitals, typically of
the $\sigma^*$ character \cite{Langhoff1984,Piancastelli19991}.

A renewed interest in studying SR's has been promoted by recent
development of laser interferometric techniques which allowed to
resolve atomic and molecular photoionization in time. One such
technique known as reconstruction of attosecond beating by
interference of two-photon transitions (RABBITT) allowed to measure
the photoelectron group delay in the SR region of various molecules:
N$_2$ \cite{PhysRevA.80.011404,Nandi2020,Loriot2021}, N$_2$O
\cite{PhysRevLett.117.093001}, CO$_2$ \cite{PhysRevA.102.023118}, NO
\cite{Gong2022} and CF$_4$ \cite{Ahmadi2022,Heck2021}. A similar SR
measurement in NO \cite{Driver2020} was conducted using an
attosecond angular streaking technique \cite{PhysRevA.106.033106}.
The photoelectron group delay, also known as the Wigner time delay,
was introduced into particle scattering theory
\cite{Eisenbud1948,PhysRev.98.145,PhysRev.118.349} and then extended
to various applications including photoionization (see reviews
\cite{deCarvalho200283,Deshmukh2021,Deshmukh2021a}).  In the presence
of a SR, the photoelectron propagation is naturally delayed relative
to the free space propagation. Thus the Wigner time delay acquires
large positive values in the hundred of attoseconds range (1~as =
$10^{-18}$~s).

In general, an accurate determination of the Wigner time delay
requires detailed knowledge of elastic scattering phases and
ionization amplitudes in various photoemission channels. Gaining
knowledge of all these quantities amounts to performing a so-called
complete photoemission experiment
\cite{Cherepkov2000,Holzmeier2021,Rist2021}. However, in a simple case
of an isolated SR in a strongly dominant channel, the Wigner time
delay can be expressed as the energy derivative of the photoelectron
scattering phase in this particular channel $\tau_{\rm W} = \partial
\delta_\ell/\partial E$. In this Letter, we demonstrate that the 
phase in such a case can be extracted straightforwardly from the
measured photoionization cross section. The latter is connected with
the phase by a simple formula $\sigma \propto\sin^2\delta_\ell$. We
derive this relation from the integral equation relating the
photoionization cross section with the transition $T$-matrix. The
diagonal $T$-matrix allows to express the unitary scattering
$S$-matrix and to find the elastic scattering phase. We examine
several shape resonances in the $nd\to\ve f, n=3,4$ ionization
channels of the Xe atom and the I$^-$ negative ion which demonstrate
the $\sigma(\delta)$ relation to a very high precision. Then we
examine the SR in the NO molecule and find a consistency between the
measured photoionization cross section \cite{Kosugi1992} and the
resonant Wigner time delay \cite{Holzmeier2021}.  This way the
experimental results obtained over a span of 30 years are seamlessly
bridged. The SR analysis in the \N molecule also supports our
findings.

\begin{figure}
\hs{-1cm}
\epsfxsize=8cm 
\epsffile{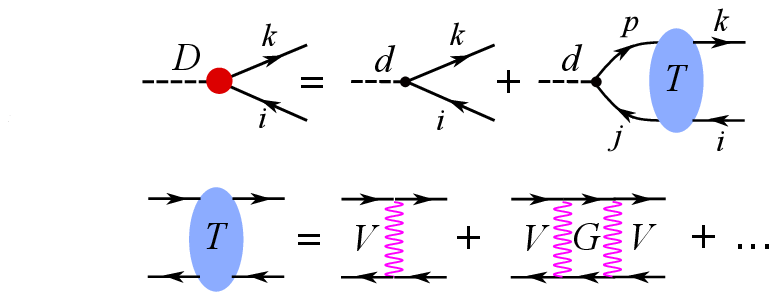}
%\epsffile{EPS/fig3g.eps}
\caption{Diagrammatic representation of the integrated dipole matrix
  element $D(k)$ (top) and the scattering $T$-matrix (bottom).  The
  following graphical symbol are in use: the straight line with an
  arrow to the right and left denotes a photoelectron and an ionic
  (hole) state, respectively. The dotted line exhibits a photon, the
  wavy line stands for the Coulomb interaction. The shaded circle and
  oval are used to represent the $D$- and $T$-matrices,
  respectively. The black dot stands for the bared dipole matrix
  element $D(k)$.
\label{Fig1}}
\end{figure}

We start our derivation by expressing the photoionization amplitude
as an integral of the dipole matrix with the transition $T$-matrix:
\be
D_i(k) = d_i(k) + \sum_j\int p^2dp \, d_j(p) G_{i,j}(p,k) T_{ij}(p,k)
\ .
\label{D-matrix}
\ee
Here the indices $i,j$ describe the residual ion and 
$p,k$ denotes the photoelectron momenta. The Green's function
\ms
\be
\label{GF}
G_{ij}(k,p) = (\ve_i+k^2-\ve_j-p^2-i\epsilon)^{-1}
\ .
\ee
accounts for the energy non-conservation in the intermediate virtual
state. We used \Eref{D-matrix} previously in convergent close-coupling
calculations of multiple atomic photoionization
\cite{BFKS02,Bray2012135}. This equation is exhibited graphically in
the top row of diagrams shown in \Fref{Fig1}. The transition
$T$-matrix is expressed via the Coulomb $V$-matrix by an infinite
sequence of diagrams displayed in the bottom row of this figure.  The
knowledge of the $T$-matrix allows to express the unitary scattering
$S$-matrix. In the case of a single-channel scattering, the said
matrix is related to the elastic scattering phase \cite{Bransden1970}:
\be
 S(k) = e^{2i\delta(k)} =  1-i2\pi k T(k,k)
= 1+2{\rm Im}\, G(k)\ T(k,k)
\ee
In the above expression, we dropped the single valued indices $i,j$
defining the ionic state in the dominant scattering channel.  The
integral equation for the ionization amplitude \eref{D-matrix} in a
single channel approximation is reduced to
\ba
\label{d-matrix}
D(k)  &=&  d(k) + \int dp \, d(p) G(p) T(p,k)
\\\nn&\approx&  d(k) {\rm Im}\, G(k) \  T(k,k)
=
\frac12 \ d(k)
\Big[
e^{2i\delta(k)}-1 
\Big]
\ea
Here we assume that the integral term in the right-hand side of
\Eref{d-matrix} is dominant over the bare term near the resonance and
the Green's function can be represented by its on-shell imaginary
part. Our numerical examples show that both assumptions are satisfied
to a high accuracy near a SR.  By squaring the modulus of the
ionization amplitude \eref{d-matrix} we arrive to the cross section
\be
\label{sigma}
\sigma=\sigma_{\rm max}\sin^2\delta(k)
\ .
\ee
Here $\sigma_{\rm max}$ is the cross-section maximum at the resonance
which corresponds to $\delta(k)=\pi/2$.  A similar expression is valid
for the SR in the electron scattering
\be
\label{Connerade}
\sigma_\ell(k) = 4\pi k^{-2}(2\ell+1)\sin^2\delta_\ell(k)
\ .
\ee
\cite{Connerade1984,Horacek2019}. The only difference is the
normalization of the cross section to the incident electron flux
rather than the cross-section maximum in \Eref{sigma}.

In our numerical demonstrations of validity of \Eref{sigma}, we use
several approximations of reducing complexity. The most
accurate photoionization calculations account for inter-channel
coupling between various atomic shells (inter-shell correlation). Such
calculations are performed using the random phase approximation (RPA)
\cite{A90}. In lesser accurate calculations, we switch off the
inter-shell correlation and evaluate the ionization amplitude as a
radial integral between the bound and continuum wave functions found
in the Hartree-Fock (HF) approximation \cite{CCR76,CCR79}.  By
observing a close agreement between the RPA and HF photoionization
cross sections we ensure the SR is indeed a single channel
phenomenon. Finally, we evaluate the cross section from the elastic
scattering phase $\delta_\ell$ found in the HF approximation. For
neutral atomic targets, we subtract the long-range Coulomb phase and
use the phase difference $\varDelta_\ell=\delta_\ell-\sigma_\ell$. It
appears that the smooth Coulomb phase plays insignificant role in the
SR formation.

\begin{figure}[t]
\bp{10cm}
\hs{-1.cm}
\epsfxsize=5.5cm 
\epsffile{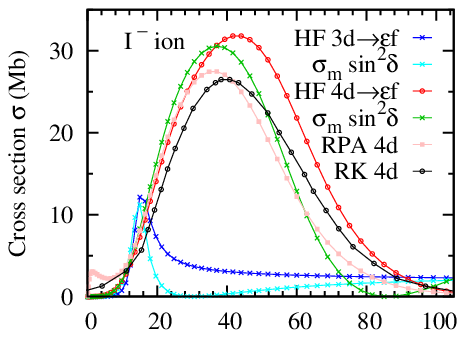}
%\epsffile{I-/CROSS.eps}
\hs{-1.5cm}
\epsfxsize=5.5cm 
\epsffile{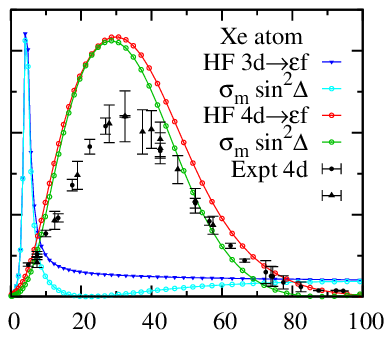}
%\epsffile{Xe/CROSS.eps}
\ms

\hs{-1.1cm}
\epsfxsize=5.5cm 
\epsffile{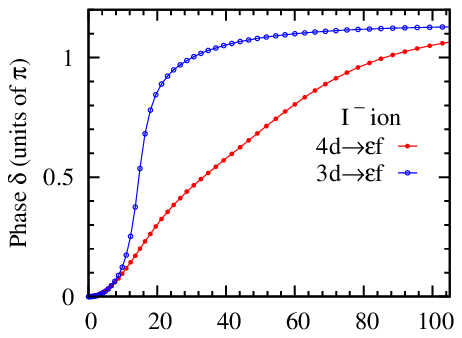}
%\epsffile{I-/DELTA.eps}
\hs{-1.5cm}
\epsfxsize=5.5cm 
\epsffile{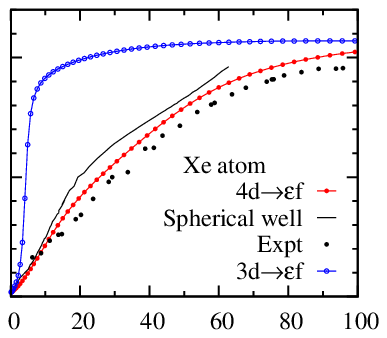}
%\epsffile{Xe/DELTA.eps}
\ms

\hs{-1.1cm}
\epsfxsize=5.5cm 
\epsffile{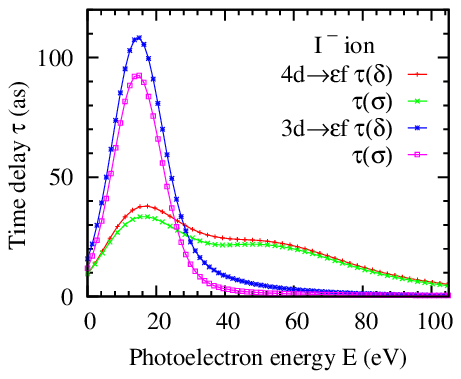}
%\epsffile{I-/TAU.eps}
\hs{-1.5cm}
\epsfxsize=5.5cm 
\epsffile{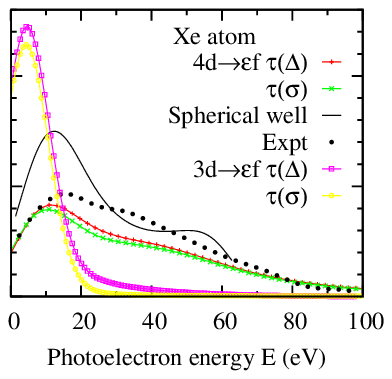}
%\epsffile{Xe/TAU.eps}

\ep

\caption{Top: photoionization cross sections of the $nd$ shells of
  I$^-$ (left) and Xe (right). The HF cross sections in the dominant
  $nd\to\ve f$ channels are compared with the cross sections derived
  from the corresponding HF phases using \Eref{sigma}.  Also shown are
  the RPA calculation for the whole $4d$ shell of I$^-$ correlated
  with the $5s$ and $5p$ shells. A similar $4d$ RRPA calculation by
  \citet{Radojevic1992} is marked RK.  The $4d$ cross section of Xe is
  compared with the experimental data
  \cite{Kammerling1989,PhysRevA.39.3902}. Bottom: time delay in the
  $nd\to\ve f,n=3,4$ channels of I$^-$ (left) and Xe (right) as
  calculated from the corresponding scattering phases and the
  photoionization cross sections. In Xe, the scattering phase and time
  delay are also  derived from the experimental
  cross-sections (dots) and the spherical well model
  \cite{Connerade1986} (black solid lines).
\ms\ms
\label{Fig2}}
\end{figure}

We start our numerical demonstrations with the SR in the $nd\to\ve f,
n=3,4$ ionization channels of the I$^-$ negative ion. We use the ionic
target to eliminate the long range Coulomb potential which would
otherwise dominate the non-resonant Wigner time delay near the
threshold.  The top left panel of \Fref{Fig2} displays the
photoionization cross sections of the $3d$ and $4d$ shells of I$^-$
calculated in HF the approximation as well as derived from the
corresponding elastic scattering phases using \Eref{sigma}. For
comparison we display the RPA calculation for the whole $4d$ shell
cross-section correlated with the $5s$ and $5p$ shells.  The
relativistic RPA (RRPA) calculation for the $4d$ shell
\cite{Radojevic1992} is also shown. We observe a close proximity of
the RPA and HF cross sections which differ only marginally from the
phase derived cross sections. The relativistic effects are also
insignificant here. These observations support our assumption that the
SR is largely a single-channel phenomenon and the cross section is
derived mostly from the elastic scattering phase in a given partial
wave. The bottom left panel of \Fref{Fig2} displays the time delay in
the $nd\to\ve f,n=3,4$ ionization channels. The two sets of
differently styled curves exhibit the time delay in each channel as
obtained by energy differentiation of the corresponding elastic
scattering phase $\tau(\delta)$ and as obtained from the
photoionization cross section $\tau(\sigma)$ using \Eref{sigma}. The
two methods of time delay determination produce very close results.

A similar data for Xe are presented in the right set of panels in
\Fref{Fig2}. The calculated $4d$ cross sections are compared with the
experimental data \cite{Kammerling1989,PhysRevA.39.3902}. For the
resonant time delay calculations, the effect of the Coulomb field is
removed by using the reduced HF phases $\varDelta_\ell =
\delta_\ell-\sigma_\ell$ where the analytically known Coulomb phases
$\sigma_\ell$ \cite{Barata2011} are subtracted. Both the phase derived
$\tau(\Delta)$ and cross section derived $\tau(\sigma)$ time delays
are even closer in Xe than in the case of I$^-$ as correlation effects
are weakened in Xe by the Coulomb field of the ion remainder.

In both considered targets, I$^-$ and Xe, the SR position depends very
strongly on the depth of the $nd$ hole.  The Coulomb attractive
potential acting on the departing photoelectron is stronger for a
deeper $3d$ hole which is screened less by outer atomic shells. Such
an un-screened Coulomb potential counters the repulsive centrifugal
potential more efficiently. As the result, the lower energy
photoelectrons are trapped into the SR. This effect is somewhat
stronger in the neutral Xe atom in comparison with the negative I$^-$
ion. The photoelectron phase variation in the SR is close to one unit
of $\pi$. When this variation occurs inside a narrow SR, the energy
derivative of the phase and the corresponding Wigner time delay
increase proportionally.

 \begin{figure}[t]
 \epsfxsize=6cm
 \epsffile{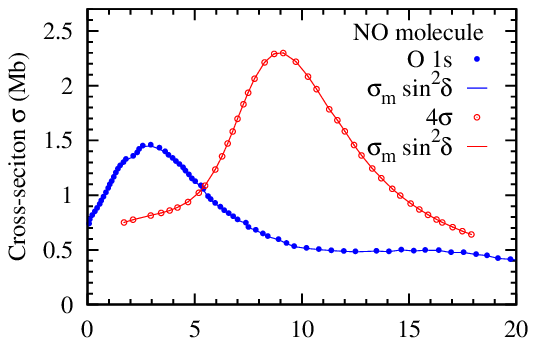}
% \epsffile{NO/CROSS.eps}
 \vs{-5mm}

 \epsfxsize=6cm 
 \epsffile{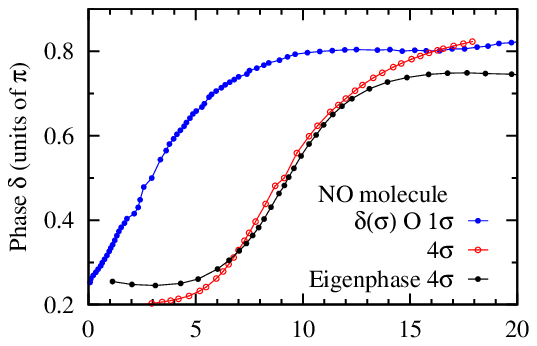}
% \epsffile{NO/DELTA.eps}
 \vs{-5mm}

 \epsfxsize=6cm 
 \epsffile{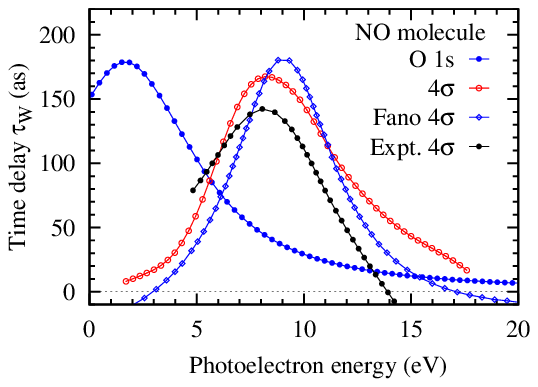}
% \epsffile{NO/TAU.eps}

 \caption{Top: photoionization cross-sections of the core O $1s$
   \cite{Kosugi1992} and valence $4\sigma$ \cite{Holzmeier2021}
   photoionization of NO. Middle: the photoelectron phases
   $\delta(\sigma)$ derived from the cross-sections exhibited in the
   top pane. The $4\sigma\to k\sigma$ eigenphase from \cite{Holzmeier2021} is
   shown for comparison. Bottom: the Wigner time delay $\tau_{\rm W}$
   obtained by energy differentiation of the phases derived from the
   corresponding cross-sections. The $\tau(\sigma)$ time delay is
   compared  with the Fano formula delays calculated and
   measured in \cite{Holzmeier2021}. 
 \label{Fig3}}
\ms
 \end{figure}

\citet{Connerade1984} applied a simple spherical well model to
describe the SR in the $4d$ photoionization of Xe and neighboring
atoms. In this mdel, the photoelectron phase in the $f$-wave is
expressed analytically via the spherical Bessel functions
\be
\label{Bessel}
\tan\delta_3 = 
{zj_3(z')j_2(z)-z'j_3(z)j_2(z')
\over
zj_3(z')j_{-3}(z)+z'j_{-4}(z)j_2(z')}
\ .
\ee
Here 
$z = ak$ and $z' =a\sqrt{k^2+2D}$
are functions of the photoelectron momentum $k$, depth $D$ and radius
$a$ of the spherical potential well. Thus found phase $\delta_3$ is
fed into \Eref{Connerade} with $\ell=3$ to find the cross-section
which is displayed in the bottom panel of Fig.~1 of
\cite{Connerade1984}. We retrofit this cross section with
\Eref{Connerade} and display thus extracted phase and time delay in
the middle and bottom right panels of \Fref{Fig2}. The time delay in
this model is markedly different from our calculations and the time
delay obtained by feeding the experimental data
\cite{Kammerling1989,PhysRevA.39.3902} into \Eref{sigma}. More
notably, the spherical well model fails completely for the SR in the
$3d$ photoionization. This indicates a much more complicated structure
of this SR.

Next, we apply our analysis to the NO molecule. Here, the SR occurs
because an unoccupied unti-bonding $6\sigma(\sigma^*)$ orbital appears
at a positive energy and merges with the $k\sigma$ final state
continua. In the meantime, an anti-bonding $2\pi(\pi^*)$ orbital falls
into the discrete spectrum and manifests itself as an isolated peak in
the photoabsorption cross section. Due to this mechanism, the $\sigma^*$
resonance is expected to be similar both in the core and valence shell
ionization. We demonstrate this effect in \Fref{Fig3} where we compare
the oxygen $1s$ \cite{Kosugi1992} and the valence $4\sigma$
\cite{Holzmeier2021} photoionization of NO. The corresponding
photoionization cross-sections are displayed in the top panel of the
figure. The absolute $4\sigma$ photoionization cross-section is read
from Fig.~1 of \cite{Holzmeier2021}. The relative O $1s$ cross-section
is read from Fig.~2 of \cite{Kosugi1992} and scaled arbitrarily. In
the middle panel of \Fref{Fig3} we display the photoelectron
scattering phases $\delta(\sigma)$ extracted from the photoionization
cross-sections exhibited in the top panel. For the valence $4\sigma$
ionization, we make a comparison with the photoelectron scattering
eigenphase exhibited in Fig.~5 of \cite{Holzmeier2021}. The latter
phase is shifted vertically to match the cross-section derived
phase. This shift does not affect the time delay which is expressed
via the phase derivative. The phase comparison shows their rather
similar slopes which are translated into similar time delays displayed
in the bottom panel of \Fref{Fig3}. In the case of the valence
$4\sigma$ ionization, the time delay compares very closely with the
Fano derived delays obtained from the calculated and measured data in
\cite{Holzmeier2021}. These observations support the validity of the
phase and time delay extraction from the corresponding photoionization
cross-sections. We also observe a rather similar phase variation and
time delay in the core and valence shell photoionization. This is in a
sharp contrast to the atomic case illustrated in \Fref{Fig2}. Such a
profound difference is explained by different mechanisms of the SR
formation. In atoms, it is a competition of the attractive Coulomb and
repulsive centrifugal potentials that leads to trapping the
photoelectron in a SR. In molecules, it is the trapping the
photoelectron into a vacant anti-bonding $\sigma^*$ orbital which is
rather insensitive to the photoelectron birth orbital. The only
difference is an insignificant shift of the resonance energy which is
marginal on the scale of the vastly different core and valence shell
ionization potentials.

 \begin{figure}[t]
 \epsfxsize=6cm
 \epsffile{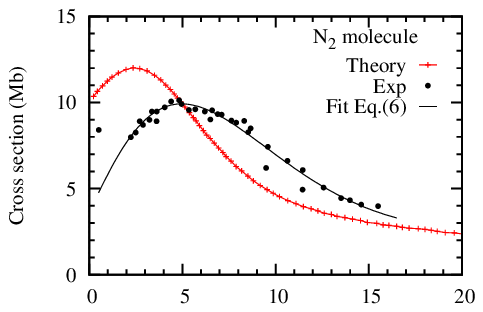}
% \epsffile{NO/PLOTN.eps}
 \vs{-6mm}

 \epsfxsize=6cm 
 \epsffile{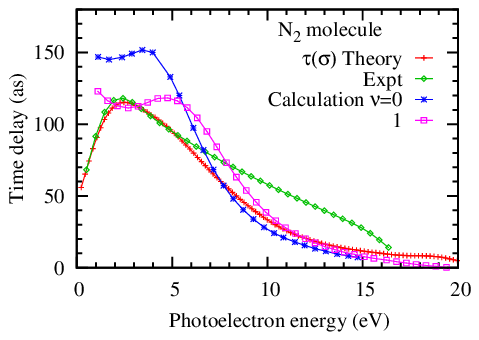}
% \epsffile{NO/PLOTNt.eps}

 \caption{Top: Photoionization cross-sections of the N$_2$ molecule as
   calculated in \cite{Nandi2020} and measured in
   \cite{Hamnett1976,Plummer1977}. The experimental cross section is
   fitted with the spherical well ansatz \eref{Connerade}.  Bottom:
   The time delay derived from the calculated and measured
   cross-sections are compared with direct calculations of \cite{Nandi2020}
for the two lowest vibrational states $\nu=0,1$ of the final
$X\,^2\Sigma^+_g$ ionic state.
 \label{Fig4}}
\ms
 \end{figure}

Finally, we derive the Wigner time delay from the 
cross-section   of the valence $3\sigma$ photoionization of the
\N molecule. Here the $(3\sigma_g^{-1})X\,^2\Sigma^+_g$ channel
contains a $\sigma\to\sigma^*$ shape resonance merging with the
$3\sigma_g\to k\sigma_u$ continuum.
For our analysis, we take the measured and calculated cross-sections
displayed in Fig.~3B of \cite{Nandi2020} and re-plot them in the top
frame of \Fref{Fig4}.  There is an insignificant energy shift between
the calculation and experiment due to the simplicity of the used
theoretical model. The experimental data points are scattered and an
analytic fit with \Eref{Connerade} is applied to feed them, along with
the calculated cross section, to \Eref{sigma}.  Thus derived phases
are converted to the Wigner time delay by energy differentiation and
the results are displayed in the bottom frame of \Fref{Fig4}. These
results are compared with the time delays calculated in
\cite{Nandi2020} for the two lowest vibrational states $\nu=0,1$ of
the final $X\,^2\Sigma^+_g$ ionic state which contains the SR. While
the fine details of the cross-section derived and directly calculated
time delays differ, the overall shape and magnitude of both sets are
quite similar.

In conclusion, we derive and test a fundamental relation between the
cross section and the time delay in the region of a shape resonance in
photoionization of atoms and molecules.  While this relation is
natural in electron scattering, it is demonstrated and rigorously
proven in photoionization for the first time.  This relation signifies
an intimate link of photoionization and electron scattering processes
which was demonstrated previously in multiple atomic photoionization
\cite{BFKS02,Bray2012135}. We support our findings by considering
several examples of atomic SR's in the $nd$ shells of Xe and I$^-$ and
molecular SR's in the $n\sigma$ shells of \N and NO. In the latter
molecule, the SR in the core O $1s$ photoionization is also
considered. In the atomic cases, the $\delta(\sigma)$ scattering phase
produces the time delay which is almost identical with the directly
calculated value. In molecules, small differences exist between the
two sets but, generally, they agree remarkably well for such a simple
model that we offer. The importance of the present findings is that
they help to link a large data set of synchrotron measurements (see
e.g. \cite{Gallagher1988}) with the recent laser based interferometric
experiments. This link offers a rigid test that allows to examine the
consistency of the two sets of data. Another important observation is
how the time delay varies when the depth of the atomic or molecular
hole state changes. In atoms, the time delay grows for inner shells in
comparison with their valence counterparts. This finding supports the
SR model in which the Coulomb field of the ionic core counterbalances
the centrifugal potential in a large $\ell$ partial wave. In
molecules, another competing explanation is more relevant in which the
SR occurs due to a trapping of the photoelectron in a non-bonding
vacant orbital. Such a trapping is rather insensitive to the birth
place of the photoelectron.

\paragraph*{Acknowledgment:}  The authors thank James
  Cryan and Taran Driver for many stimulating discussions.

%\np~~~\np

%\bibliography{references,mypapers,areferences,dreferences,hreferences,H2references,ureferences,wreferences,Wreferences,xreferences}

\end{document}